# The New Threats of Information Hiding: the Road Ahead

K. Cabaj[1], L. Caviglione[2], W. Mazurczyk[1], S. Wendzel[3], A. Woodward[4], S. Zander[5]
[1] Warsaw University of Technology, Poland
[2] National Research Council of Italy (CNR), Italy
[3] Worms University of Applied Sciences, Germany
[4] University of Surrey, UK
[5] Murdoch University, Australia

**Abstract**
Compared to cryptography, steganography is a less discussed domain. However, there is a recent trend of exploiting various information hiding techniques to empower malware, for instance to bypass security frameworks of mobile devices or to exfiltrate sensitive data. This is mostly due to the need to counteract increasingly sophisticated security mechanisms, such as code analysis, runtime countermeasures, or real-time traffic inspection tools. In this perspective, this paper presents malware exploiting information hiding in a broad sense, i.e., it does not focus on classical covert channels, but also discusses other camouflage techniques. Differently from other works, this paper solely focuses on real-world threats observed in the 2011 – 2017 timeframe. The observation indicates a growing number of malware equipped with some form of data hiding capabilities and a lack of effective and universal countermeasures.

## 1. Introduction

The use of *information hiding* techniques, often generically referred to as *steganography*, to commit cyber-attacks or crimes has received relatively little attention in the academic literature or the media. When mentioned, steganography is typically discussed only in the context of covert communication between extremist individuals or groups [1]. Even then, some argue that there is little or no evidence that steganography is in use. While large-scale surveys found no conclusive traces of the use of data hiding, some authors warn against concluding that it is not in use [2]. Only recently there have been some signs that things are starting to change. Reports from McAfee [3] and Kaspersky [4] recognized the role that information hiding has in current malicious software and that it is highly likely to gain additional importance in the future. Furthermore, because of the sensitivity of the subject in many cases organizations are reluctant to report the detected use of steganography to the public [5].

Historically, *cryptography* has been a more widely discussed topic, especially for law enforcement purposes. While in the past the mere existence of encrypted communications and data may have raised suspicions, today it is a frequent scenario. For example, malware using encrypted communications for Command and Control (C&C) purposes might previously have stood out from regular network traffic, but now it is effectively hidden within the "background noise" of routinely encrypted data exchanged in the network. Nevertheless, encrypted communications can be detected relatively easily and ancillary techniques, such as traffic analysis or metadata recovery, enable at least some intelligence to be derived from encrypted data and communications. Besides, the recovered metadata, e.g., who is communicating with whom, when and for how long, can be as important as, or even more important, than knowing the actual content.

Currently, encryption is receiving greater attention from security professionals, law enforcement, security and intelligence agencies. For example, recent advancements in understanding how malicious software encrypts its own communications could help identify and block C&C communications of botnets [6]. Unfortunately, criminals or extremists are well aware of the increased focus on encryption and are looking for other ways to make malicious software stay "under the radar", especially in the context of stealing data where triggering some form of defense must be avoided. In this vein, the most important and recent trend is to equip malware with information hiding capabilities, i.e., techniques that hide communications [7].

This paper provides an overview of information hiding techniques that can be utilized by malware. By using real-world examples, this paper showcases existing and emerging threats using different types of data hiding mechanisms, i.e., not only limited to those adopting classical covert channels. The research presented here has been performed within the Criminal Use of Information Hiding (CUIng) initiative (*cuing.org*) that has been formed with the cooperation of the Europol European Cyber Crime Centre (EC3) for gathering experts from different backgrounds with the aim of monitoring information-hiding-capable threats and propose efficient countermeasures.

The remainder of the paper is structured as follows: Section 2 introduces the general concepts of data hiding and covert channels. Section 3 analyzes existing malware using information hiding, while Section 4 discusses the road ahead from the perspective of the new challenges arising from the future use of information hiding in malware. Section 5 concludes the paper.

## 2. Covert Channels and Data Hiding

In this section, we introduce the fundamentals of generally perceived information hiding and its sub-disciplines, and how they are related to malware and the different phases of cyber-attacks. Such attacks are commonly divided into five phases [8]: (1) reconnaissance (gathering of information), (2) scanning the target, (3) gaining access to the target, (4) maintaining the access and (5) covering the tracks. However, it must be noted that information hiding techniques are mostly applied in phases 2-4 on which we focus here.

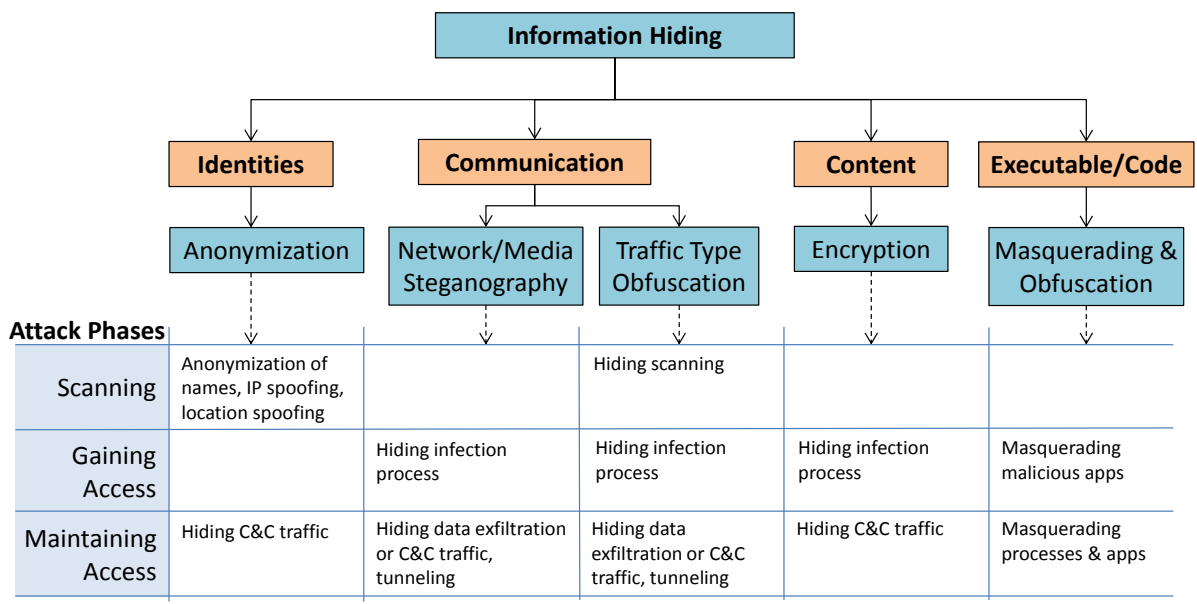

**Figure 1: Classification of hiding techniques and how they are used by malware in the different attack phases**

As depicted in Fig. 1 information hiding is a very broad term. It encompasses different "sub-disciplines" (or "domains"), which can be used by an attacker during different attack stages depending on *what* is subjected to hiding:

- *Identities*: the identities of communicating parties are hidden by anonymization techniques.
- *Communication*: the fact that a communication is taking place is hidden by steganography techniques. The very characteristics of a network conversation (e.g., a packet flow) can be concealed using traffic type obfuscation methods.
- *Content*: hiding solely the content of data but not the transmission or presence of the data itself is achieved by applying cryptographic algorithms.
- *Code*: the structure of (executable) code is hidden by (binary) code obfuscation and masquerading techniques.

First, let us discuss the most important (from this paper's perspective) data hiding methods which conceal the fact that a communication is taking place. Typically this type of information hiding is realized using some form of *steganography*.

Historically, the earliest computer steganographic methods were focused on different media types, especially digital images. As an example, several algorithms hide information within the least significant bits of color definitions of pixels within an image as the human eye cannot spot such alterations. A similar approach has been used for audio and video. The natural evolution is to hide data in network transmissions, such as in inter-arrival times of packets or in unused fields of protocol headers. Network traffic provides the advantage of a continuous data flow, which a digital media file of constant size cannot provide. When secret data is hidden in network traffic, the secret communication channel is referred to as a *network covert channel*.

In essence, network covert channels enable secret malware communications over any type of computer network, be it a local area network or the Internet. Compared to encryption, which only ensures the confidentiality of what a malware communicates, covert channels can help to keep the communication secret, e.g. to retain access to a hacked system. Moreover, control protocols can be used on top of covert channels, representing a form of C&C channel. Such control protocols allow to upload a newer version of a malware binary, to select a different encryption or covert signing scheme, to switch from one steganographic method to another or to apply dynamic overlay routing to bypass firewalls [9]. Malware can also apply network covert channels to conceal the exfiltration of organizational data over the network and to bypass firewalls by hiding data in transmissions that are not affected by its filtering policy. These goals affect especially attack phases 2 and 3 of Fig .1 (gaining and maintaining access). We point out that when referring to malware trying to communicate covertly or abusing some network service, the hacking community often uses the term *tunneling*. However, this is not accurate as tunneling hides traffic as a "by product", and actually refers to the encapsulation of network data of the same or higher layer, e.g., IPv4 as payload in an IPv6 packet.

While steganography aims to hide data inside of digital objects, two other classes of methods obfuscate information in code (*code obfuscation*) or network traffic (*traffic type obfuscation*). Obfuscation is different from steganography as the latter tries to communicate secret data in a non-noticeable manner while the former is directly visible to an analyst. Despite of their different strategies, both domains share the goal of hiding data. The goals of traffic type and code obfuscation affect phase 1 (scanning) but mainly phases 2 and 3 of Fig.1 (gaining and maintaining access).

*Anonymization* provides means of communication without revealing private attributes of the communicating peers, such as their names, their IP addresses or geographical locations. In contrast to steganography, anonymization relies upon different techniques, such as spoofing of the IP address of a sender or cryptographic algorithms to fake or hide sensitive data that can be used to deduce information about who is involved in a communication. Note that, as shown in Fig. 1, cryptographic methods can be used to encrypt any kind of secret data, not just data that reveals identities. Thus, the application of cryptography is not limited to anonymity techniques. Anonymity techniques can be utilized during phases 1 (scanning) and 3 (maintaining access) while encryption (despite of its use for anonymity purposes) affects phases 2-3 (gaining and maintaining access).

### 3. Information Hiding Malware in to the Wild

In this section we present several examples of information hiding-capable malware observed in to the wild owing to the CUIng initiative. Due to space constraints, we focus only on the most representative threats observed in the 2011-2017 period.

Originally, information hiding techniques were implemented only in Advanced Persistent Threats (APTs) like Duqu, Regin or Hammertoss, which are the most sophisticated types of malware probably created with the support of nation-wide sponsors. However, as they proved their effectiveness, they are slowly becoming the de-facto standard also for the "ordinary" malware. As a demonstration, various types of currently popular threats like ransomware (e.g., TeslaCrypt, Cerber and SyncCrypt) or exploit-kits (Stego/Astrum, DNSChanger, and Sundown) use some form information hiding. This trend is reviewed in detail below.

*3.1 Malware using modifications to digital media files*

Currently, one of the most common ways for hiding data is to use digital media files as the secret carrier. The most common technique exploits digital images to: *i)* conceal malware settings or a configuration file; *ii)* provide the malware an URL from which additional components can be downloaded from; *iii)* store directly the whole malicious code. The most notable examples, in 2015, Vawtrak/Neverquest malware started utilizing steganography to hide settings in favicons, i.e., innocent-looking pictures widely available in websites. The malware extracts the least significant bits from each image's pixel in order to reconstruct a previously embedded URL for downloading its configuration file. A similar approach has been used by Zbot malware, which downloaded an innocently-looking JPEG image on the infected system containing its configuration data appended at the end of the image. Finally, Lurk and Stegoloader used the least significant bit of a digital image (a BMP and PNG, respectively) to retrieve an encrypted URL for downloading additional software components.

More recently, we also observed the use of information hiding techniques for *malvertising* attacks as evidenced by the AdGholas malware. AdGholas avoids detection by using steganography for hiding encrypted JavaScript code in images, text and HTML code. Lastly, at the end of 2016 large-scale attacks related to the online e-commerce platform Magento revealed the usage of image steganography to conceal details of payment cards. Specifically, once the platform was infected, the malware collected payment details and hid them inside images of real products available on the infected e-commerce site. By downloading such modified images, the attacker could easily exfiltrate the stolen data.

*3.2 Malware posing as other legitimate applications or mimicking their traffic behavior*
In this case, the malware relies on the mimicry of legitimate programs and/or their communications. A paradigmatic example is a variant of Android/Twitoor.A, a malware spreading by SMS or via malicious URLs. The malware impersonates a porn player app or an MMS application but without having their functionality, eventually tricking the user to install them and spread the infection. Another application, Irongate is the first notable example designed to operate in industrial control systems scenarios. One of the most important features is its ability to record several seconds of ordinary, legitimate traffic from a programmable logic controller and then use it as a smokescreen (i.e. the malicious commands are masked using legitimate ones) when sending intentionally modified data back. Such operation allowed the attacker to alter a controlled process without raising any security alerts.

Another example is Fakem RAT that made its C&C traffic look like MSN and Yahoo! Messenger or HTTP conversations. At the beginning of 2017 Carbanak/Anunak demonstrated its ability in abusing Google cloud-based services to set-up a covert channel for C&C purposes. In this case, a unique Google Sheets spreadsheet is dynamically created in order to manage each infected victim. The use of a Google service granted attackers the ability to stay under the radar as typically such third party services are not blocked in the enterprise network and considered safe. Another example using a similar technique includes a new version of the SpyNote Trojan, which was disguised as a legitimate Netflix application. Once installed, it allowed the attacker to execute different actions, such as copy a user's files, view a user's contacts, and eavesdrop a user's communication. More recently, a technique called *domain fronting* is gaining a lot of attention especially among APT-related groups. Put briefly, it is used to mask the true destination of a connection by mimicking legitimate traffic to an innocent destination. A successful implementation exploits HTTPS traffic to communicate with an infected host and making the traffic look like a Google search. Instead the traffic is produced by a connection exchanging data with the attacker.

*3.3 Information hiding in ransomware*
First experiments of this kind were discovered at the beginning of 2016 when TeslaCrypt has been spread using the Neutrino exploit kit. The infection process is as follows. Neutrino initially redirects users to a malicious landing page crafted for discovering the vulnerabilities of the victim to deliver the most appropriate exploit. If the vulnerability is successfully exploited, a downloader is executed. To gather data, it contacts a server, which responds with an HTTP 404 error page embedding C&C commands in the HTML comments tag. Next, in mid-2016, Cerber has been identified as one of the macro-type malware-delivered ransomware across a variety of cloud-based file-sharing applications. To spread the infection, Cerber uses a decoy document which, when opened, loads a malicious macro-code that downloaded a JPEG file to the targeted machine. Inside this benign-looking image was the steganographically embedded malicious executable. Finally, in August 2017 a similar technique has been discovered in the SyncCrypt ransomware. Infected emails contained WSF (Windows Script File) attachments posing as court orders. If opened, a malicious code downloaded a digital image containing the core components of SyncCrypt.

*3.4 Information hiding in exploit kits*
This is one of the most recent trends. In this case, information hiding methods became so popular among cybercriminals that they are incorporated within exploit kits to allow developers with little or no programming skills to create, customize and distribute malware. The first example showing this feature is the Stegano/Astrum exploit kit, which has been used at the end of 2016 as part of a huge malvertising campaign. Its developers were able to embed malicious code within banner ads by modifying the color space of the used PNG image (i.e., the alpha channel). Then, the browsers of users viewing an infected ad parse an injected JavaScript code extracting the malicious code and redirecting users to the exploit kit landing page. On the landing page the infection with different breeds of malware is performed, typically by using several Flash vulnerabilities.

In 2016 another type of exploit kit relying upon malvertising has been identified. DNSChanger hides an AES encryption key within an innocent looking ad to decrypt the network traffic generated by the exploit kit. The scope of DNSChanger is to launch brute-force attacks against the network routers to take control over the victim's network and inject ads in all exchanged traffic. While Stegano/Astrum and DNSChanger are niche products, the Sundown exploit kit is one of the major players in the exploit kit market that introduced data hiding methods to the wider cybercriminal audience. In particular, Sundown uses steganography in two ways: *i)* to covertly exfiltrate information stolen from the infected system in PNG files which are uploaded to an Imgur album where cybercriminals can access them undisturbed (see, the CryLocker ransomware campaign); *ii)* to hide the exploit code delivered to the victims.

**4. The Road Ahead**
As hinted above, in the past information hiding has been mainly associated with the world of espionage or extremism and trends were driven by developments in this area. However, in recent years we have experienced a massive growth in cybercrime and due to the lucrativeness of this "business" this tendency is likely to continue [3]. In the area of cybercrime we see the following main developments: increased stealth, commoditization of malware, and exploitation of Internet of Things (IoT) devices. Cybercriminals will put a higher emphasis on making it harder to detect and trace back malware to the origin and this will be a main driver for the increased future use of information hiding.

Since a main goal of malware developers is to be one step ahead, they will always try to improve their information hiding techniques. One avenue is to utilize better digital media steganography algorithms. Improved algorithms, which are harder to detect and eliminate than the currently used ones, are already available and known among academics (e.g. F5 [13], HUGO [14]). Another strategy is to hide in new services or protocols. Services and protocols, such as Skype/VoIP [15], BitTorrent, SCTP [16], can be targeted resulting in a "needle in a hay stack" problem when it comes to detecting covert communication among a large number of similar connections.

Another future direction is to exploit the currently ongoing IPv4 to IPv6 transition. Malware can take advantage of misconfigured nodes or hosts with IPv4-only stacks unable to process IPv6 malicious traffic. Malware also increasingly exploits the diffusion of HTTPS by hiding in HTTPS or TLS traffic, which cannot be easily inspected (see, e.g., [10] which claims that over one third of malware already uses HTTPS).

Botnets will remain an important tool for cybercriminals for various purposes, such as managing DDoS attacks or sending spam emails. Since bots can be relatively easily identified by observing the C&C traffic, masquerading this traffic is very important. While most existing approaches are simple, for example, C&C protocols hide in HTTP, IRC or DNS, academic research recently demonstrated how to completely transform a C&C protocol to mimic another innocuous protocol [18]. Future botnets may utilize overlay networks that use only steganographic methods to communicate (stego-botnets) [17].

The DNS protocol is a natural choice as cover for C&C traffic or for data exfiltration as it cannot be blocked. Developing stealthier covert channels on top of DNS and developing the countermeasures to detect these is an ongoing arms race [19], which could become even more interesting once DNSSEC will be more widely deployed.

Another ongoing trend is that attacks will more and more target the ever-increasing number of IoT devices, such as networked sensors, CCTV cameras, smart TVs and DVRs, smart home and building appliances and industrial control systems, rather than the traditional devices, such as PCs. In many cases IoT devices are soft targets as their limited processing capabilities limit the implemented security mechanisms and even worse the low cost of many of these devices means that security is often an afterthought for manufacturers. Attacks on IoT devices also allow user profiling, and maliciously interfering with the physical world. Moreover, malware can utilize the IoT to hide secret data [11], for instance, an attacker can secretly store data in unused registers of IoT devices or by slightly modifying actuator states [12].

Currently deployed steganography methods are often simple. The main reason for this is that current protection solutions, such as intrusion detection systems (IDSs), hardly detect any form of steganography in practice and thus, malware developers are not forced to apply more sophisticated steganography. Nevertheless, recent threats often merge simple traffic covert channel techniques with memory-resident or file-less implementations to make them stealthier and able to cover the tracks on the infected host, e.g., in the filesystem.

However, data leakage protection (DLP) solutions increasingly aim to detect steganographic transmissions and this will force malware authors to improve the covertness of their data leakage techniques. That said, cybercriminals will increasingly choose off-the-shelf malware rather than develop custom malware, which would require higher investment. Once more advanced steganography finds its way into off-the-shelf malware products, such as exploit kits, it would become widely used at relatively little extra cost to the cybercriminals.

In the future, malware de-obfuscation and steganography analysis must be done in a more systematic and efficient way when the volume of more sophisticated malware increases. Frameworks for distributed and automated malware analysis like MASS [20] could be a suitable approach for handling large volumes of malware samples retrieved from honeynets.

## 5. Conclusions and Future Work

In this paper we presented the knowledge gathered within the CUIng initiative on threats using information hiding. As shown, a very important lesson learned is about the efficiency achieved by modern malware to remain covert for a long time. Even if steganographic techniques are not the only ingredient for this success, the ability of creating and exploiting covert channels for C&C and exfiltration purposes surely plays a role (see, e.g., Regin which acted covered from 2008 to 2014). This fact is exacerbated by a worrying lack of techniques for detecting information-hiding-capable threats, especially in the domain of IoT/automation equipment. A possible cause is the poor generalizability of the process of detecting information hiding; many detection techniques are tightly coupled with specific hiding methods, the cover they use, the scenario in which they are used and the technology on which they depend. Since creating new hiding methods by applying known techniques to new protocols, scenarios and technology is relatively easy, countermeasures are always at least one step behind. Therefore, the industry and the academia should focus on the investigation and the development of new and general tools or add-ons for the most common network security solutions. A possible idea to avoid a plethora of plug-ins or standalone solution for coping with malware using information hiding is the use of new more general indicators, such as general patterns used by different hiding techniques or energy consumption.

As shown, information hiding increases the complexity of the problem space to be faced when addressing cybersecurity. Organized initiatives like CUIng can be the incubator where a long-term cure for information hiding malware is developed, since modern cyber-threats require a "multidisciplinary" approach relying on the collaboration of many experts from the industry, academia, and law enforcement agencies.